# Controlled Growth, Patterning and Placement of Carbon Nanotube Thin Films

V. K. Sangwan[a,b]*[#], V. W. Ballarotto[b], D. R. Hines[b], M. S. Fuhrer[a], and E. D. Williams[a,b]

[a] *Center for Nanophysics and Advanced Materials, University of Maryland, College Park, MD*
[b] *Laboratory for Physical Science, College Park, MD*

**Abstract**

Controlled growth, patterning and placement of carbon nanotube (CNT) thin films for electronic applications are demonstrated. The density of CNT films is controlled by optimizing the feed gas composition as well as the concentration of growth catalyst in a chemical vapor deposition process. Densities of CNTs ranging from 0.02 CNTs/$\mu m^2$ to 1.29 CNTs/$\mu m^2$ are obtained. The resulting pristine CNT thin films are then successfully patterned using either pre-growth or post-growth techniques. By developing a layered photoresist process that is compatible with ferric nitrate catalyst, significant improvements over popular pre-growth patterning methods are obtained. Limitations of traditional post-growth patterning methods are circumvented by selective transfer printing of CNTs with either thermoplastic or metallic stamps. Resulting as-grown patterns of CNT thin films have edge roughness (< 1 μm) and resolution (< 5 μm) comparable to standard photolithography. Bottom gate CNT thin film devices are fabricated with field-effect mobilities up to 20 $cm^2$/Vs and on/off ratios of the order of $10^3$. The patterning and transfer printing methods discussed here have a potential to be generalized to include other nanomaterials in new device configurations.

* Current address: Material Science and Engineering, Northwestern University, IL 60208

# email: v-sangwan@northwestern.edu

# 1. Introduction

Excellent electrical properties of carbon nanotube (CNT) thin films can only be fully realized by controlling the density of CNTs, patterning individual device components and assembling them in desired device architecture. Incorporation of CNT thin films in electronic devices brings certain requirements on patterning methods. First of all, a patterning method needs to be compatible with CNT thin film growth process as well as the subsequent device fabrication steps. Secondly, patterned CNT thin films need to have desired resolution and edge roughness for large scale integration. Thirdly, as-grown pristine CNT thin films (i.e. CNTs are not exposed to solvents or polymer resists) are preferred for some applications. Standard patterning methods meet the above requirements with varying degree. Therefore, there is a need for versatile method to enable controlled placement of CNTs onto different device substrates and have the capability to produce large area arrays.

Random networks of CNTs are particularly useful for large scale integration due to the reproducibility of ensemble-averaged electrical response. Since the electrical response of a random network of CNTs is determined by percolation of both metallic and semiconducting CNTs, the density of the CNTs can be used to determine the electrical behavior of a CNT thin film [1-3]. It is possible to realize different CNT device components by controlling the density of the CNT thin film so the desired electrical response is achieved. However, fabrication of different device elements requires different patterning approaches.

CNTs can be patterned either by patterning the catalyst before growth or by selective etching of CNTs after growth. Most of the reported methods to pattern catalysts, such as photolithography [4, 5], electron beam lithography [6], shadow mask [7], and soft-lithography [8] are either optimized for individual CNTs or have limited compatibility with catalyst processing. For example, ferritin solution in DI water is compatible with several photoresists (Futurrex Inc); whereas, ferric nitrate solution in organic solvents is incompatible with the same photoresists. On the other hand, CNTs patterned by reactive ion etching (RIE) [9, 10] and $CO_2$ snow jet treatment [2] are exposed to resist processing. Therefore, our goal is to achieve resist- and catalyst processing compatibility and also avoid any addition resist processing on CNTs after the growth.

In this paper, we report on various methods to achieve controlled growth, patterning, and placement of CNT thin films for electronic applications. Though high quality devices have been reported using CNTs thin films [2, 9], there has been only one systematic attempt to control and characterize the density of CNT thin films to our knowledge [7]. We used chemical vapor deposition (CVD) to grow CNT thin films in conjunction with pre-growth and post-growth methods to control the density. The resulting CNT thin film arrays, which are pristine, are placed in a controlled manner onto plastic substrates with transfer printing techniques. Thus, high-quality bottom gate CNT thin film transistors (TFTs) on plastic substrates are obtained with field-effect mobility up to 20 $cm^2$/Vs and on/off ratios of approximately $10^3$.

## 2. Experimental Methods
### 2.a Controlled growth of CNT thin films

The growth of CNT films was controlled by two methods: by varying the growth conditions and by varying the concentration of the catalyst. In the first method, the density of the CNTs thin films was varied by changing relative composition of the feed gases $H_2$ and $CH_4$. Catalyst-soaked chips (70 μg/ml ferric nitrate solution in IPA) were loaded onto a quartz boat and set in a single zone tube furnace (Lindberg/BlueM) with a 1 inch quartz glass tube. CNT thin films were grown at 850 °C for 10 min by flowing $H_2$ and $CH_4$, after putting the samples through several ramping and soaking steps. Several samples were grown at varying composition of $CH_4$ (from 20 % to 90 %) while keeping the total gas flow constant at 2000 ml/min at 1 atm pressure. CNT thin films were then imaged in field emission scanning electron microscope (FE-SEM) and analyzed by the method described below.

In the second growth method, the density of CNT thin films was controlled by varying the concentration of the ferric nitrate solution. Ferric nitrate solutions of ten different concentrations (varying from 2 μg/ml to 133 μg/ml) were prepared and then, dispersed on different $SiO_2$ chips. CNT thin films were again grown at 850 $^o$C for 10 min. The gas flow rate of $CH_4$, $H_2$ and $C_2H_4$ was kept at 1900 ml/min, 1300 ml/min and 8 ml/min at 1 atm pressure, respectively. We note that a small amount of $C_2H_4$ was found to be critical in this process; absence of $C_2H_4$ resulted in sparse CNT networks even with highly concentrated catalyst. Note that presence of $C_2H_4$ in the first growth method resulted in dense and uncontrolled CNT thin films, and therefore, no $C_2H_4$ was used in the first growth method.

The density of CNTs was determined by counting the number of CNTs in field-emission scanning electron microscope (FE-SEM) images. CNTs were counted in 40 μm x 60 μm rectangles (magnification = 5 kX) for relatively sparse CNTs thin films ($\leq$ 0.57 CNTs/μm$^2$) and in 12 μm x 16 μm rectangles (magnification = 20 kX) for relatively dense CNT thin films ($>$ 0.57 CNTs/μm$^2$). Density of CNTs was averaged from two different places on the substrates. Number of CNTs crossing rectangular boundaries was divided by two to account for their partial contribution to CNT density.

**2.b Patterning CNT thin films**

The CNT thin films can be patterned in two ways. In the pre-growth method, the growth of CNTs is restricted in certain regions by patterning the catalyst particles. Pre-growth patterning ensures pristine CNT thin films, but the edge roughness of patterns is limited by length of CNTs. We note that the pre-patterned resist has to be compatible with chemical processing involved in deposition of catalyst. In the post-growth method, CNT thin films are selectively removed after growth with a transfer printing method. We have previously shown that CNTs can be transferred from $SiO_2$ to thermoplastics (polyterephthalate (PET)) [10, 11] and metallic substrates (Au and Pd) [12]. This means that pristine CNTs can be patterned by printing onto stamps made from these materials, thus a transfer printing method using stamps of PET and Au was developed.

Popular pre-growth and post-growth methods were also used for comparison with patterning methods developed here. In pre-growth method, the catalyst is patterned by photolithography and CNTs are grown after the lift-off, (Fig. 1(a)). In post-growth method, a photoresist is patterned on CNTs and exposed CNTs are etched by RIE in an $O_2$ plasma, (Fig. 1(b)). Fig. 1(c) shows an SEM image of the edge of a CNT pattern.

**2.b.1 Pre-growth patterning of CNT thin films**

A triple-layered resist processing is developed to ensure compatibility with ferric nitrate catalyst deposition. This is done by first patterning positive photoresist mesas on a 500 nm thick thermally oxidized Si substrate in the regions where patterned CNT thin films are desired (Fig. 2(a)). Then a layer of poly-methylmethacrylate (PMMA) (A4, MicroChem Corp.) was spin-coated and hard baked (180 $^o$C for 5 min) on the $SiO_2$/Si substrate, followed by a negative photoresist layer (NR7-1500 PY) that was spin-coated, baked, exposed and developed. Thus, windows were opened in the negative photoresist to expose the underlying PMMA layer. The same photomask can be used to pattern positive photoresist mesas as well as negative photoresist windows. The PMMA layer is removed by RIE with an $O_2$ plasma (power = 100 W, pressure = 200 mTorr). The negative photoresist mask also gets etched by RIE, but the layer of the photoresist (1.5 μm) is much larger than the PMMA layer (200 nm). Thus, the etching time can be optimized such that PMMA layer is completely etched before the negative photoresist (45 sec for 200 nm thick PMMA layer). The samples are next exposed to UV light to make the positive photoresist soluble in developer (Fig. 2(a)). Upon removal of the positive photoresist, the ferric nitrate catalyst can be patterned through PMMA-photoresist bi-layer. Fig. 2(b) shows an SEM image of the edge of a pre-growth patterned CNT thin film.

### 2.b.2 Post-growth patterning of CNT thin films

First, we discuss the patterning method developed for PET stamps. The process is accomplished in three steps: fabrication of a master template, molding of PET stamps from the master template, and finally transfer printing CNTs using PET stamps. To make a master template from Si, standard photolithography was done on a Si substrate to create the desired patterns of photoresist. The photoresist is used as a mask to etch the exposed Si surface by RIE, using a mixture of $CHF_3$ and $O_2$ plasma (power = 175 W, pressure = 40 mTorr). The Si substrate was etched to obtain 16 μm tall mesas that have an area of 200 μm x 300 μm. The master template is used to fabricate the PET stamps. The master Si template was covered with a self-assembled monolayer (SAM) of (tridecafluoro-1,1,2,2- tetrahydrooctyl) trichlorosilane that acts as a release layer to reduce the adhesion between the Si and PET substrates (Dupont Mulinex 453/700)[13]. Then, a PET stamp was made by pressing a PET substrate between the Si template and a plane Si substrate at 600 psi, 170 $^oC$ for 15 min in a nano-imprint machine (Nanonex 2500). A single master template can be used to make several PET stamps without degradation. In the final step, the CNT growth substrates were pressed against PET stamps at 400 psi, 140 $^oC$ for 3 min (Fig. 3(a)). Upon separation, patterned arrays of CNTs are lifted from the growth substrate.

More precise and smaller CNT patterns may be desirable for dense arrays of smaller devices. This can be achieved by using Au stamps. The Au stamps were made by evaporating 500 nm thick Au film through patterned photoresist on Si Substrate. The CNT growth substrates were pressed against Au stamps at 300 psi, 90 $^oC$ for 3 min (Fig. 3(c)). After separation, patterned arrays of CNTs are removed from the growth substrate. There was no significant deformation in Au stamps at 300 psi, therefore the stamps can be used multiple times. When CNT patterns are desired in specific regions, Au stamps can be made on transparent substrates such as quartz to facilitate alignment.

**2.c Controlled placement of patterned CNT thin films**

Controlled placement of patterned CNT films can achieved by transfer printing [10, 11] when proper differential adhesion exists between the CNT/SiO$_2$ substrate and device substrate. This is ensured by using the release layer mentioned in previous section. However, the release layer has to be patterned on the transfer substrate (SiO$_2$), but not on the transferable layer (CNTs). Thus, a micro-contact printing method was developed with PDMS stamps to pattern the release layer. PDMS (Sylgard 184) stamps were molded from the same master Si template that was used to mold PET stamps. After curing the elastomer (120 $^o$C, 1 h), the raised parts of PDMS stamp are covered with release layer by gently placing over a separate Si substrate covered with release layer (Fig. 4(a)). The release layer covered PDMS stamp was then aligned over patterned CNT thin films with help of markers such that CNT thin films lie under the trenches (Fig.4(b)). Thus, SiO$_2$ surface outside the patterned CNT thin film is treated with release layer and CNTs underneath the trench remain unaffected. Using this approach, we achieved almost complete transfer of as-grown patterned CNT thin films onto PET as well as PMMA surface with controlled placement, Fig. 5(a). Fig. 5 (b) and (c) show AFM images of SiO$_2$ transfer substrate before and after transfer printing onto a PET substrate at 500 psi, 170 $^o$C for 3 min. However, resolution of this method is limited by alignment capabilities. Here alignment accuracy of less than 10 μm was achieved using a Nanonex 2500 imprinter alignment tool. Alignment resolution can be further improved by using a higher resolution objective lens.

**2.d Device fabrication**

The above methods were used to fabricate arrays of bottom gate CNT thin film transistors. The devices layers were assembled sequentially by transfer printing as described in detail in ref [10]. Briefly, first 100 nm thick Au gate electrodes were transfer printed onto PET device substrate. Then, 50 nm thick Al$_2$O$_3$ layer was deposited on gate electrodes to avoid gate leakage from CNTs. A 800 nm thick PMMA dielectric layer was spin-coated on PET substrate, followed by transfer printing of 30 nm thick source-drain Au electrodes. Finally, patterned CNT thin film was transfer printed from release layer treated $SiO_2$/Si substrate onto Au source-drain electrodes at 500 psi, 170 $^oC$ for 3 min. An array of bottom gate TFTs were fabricated with channel width (*W*) 100 μm and channel length (*L*) varying from 1 μm to 100 μm

## 3. Results

First we report on the control of CNT growth. In Fig. 6(a), the average density of CNTs from different samples is plotted against the ratio of $CH_4$ to the total gas. The concentration of ferric nitrate catalyst was 70 μg/ml. No CNTs were observed for 20% or less $CH_4$, consistent with earlier experiment with ferritin [7]. The highest density of CNTs (0.07 CNTs/$\mu m^2$) was achieved at 75% $CH_4$ for different composition of $CH_4$ considered here. No CNTs were observed for 90% $CH_4$, possibly due to poisoning of the catalyst by excessive carbonation or pyrolysis [7]. It is worth noting that the highest density obtained by this method is less than the highest density achieved by a similar growth recipe used for ferritin and sputtered Fe films in ref [7]. This method can be useful as a fine control over density of sparse CNT networks. A broader range of CNT densities was obtained by an alternative method, as discussed below.

Fig. 6(b) shows variation in the average density of CNTs as a function of concentration of the catalyst. The flow rates of $CH_4$, $H_2$ and $C_2H_4$ were 1900 ml/min, 1300 ml/min and 8 ml/min, respectively. Notice that the density of CNTs increases linearly with concentration at low concentrations, and becomes almost constant for concentrations higher than 66 μg/ml. Thus, this method can be used to produce CNTs densities over a broad range varying from 0.04 CNTs/$\mu m^2$ to 1.29 CNTs/$\mu m^2$. Fig. 6(c), (d) and (e) show FE-SEM images of CNT thin films with 0.16 CNTs/$\mu m^2$, 0.63 CNTs/$\mu m^2$ and 1.15 CNTs/$\mu m^2$ obtained from catalyst concentrations of 5 μg/ml, 20 μg/ml and 66 μg/ml, respectively. We also observed slight variation in CNT lengths around a root mean square length ~ 5 μm with CNT density, especially at lower densities.

Thus, we can use our pre-growth method of patterning the catalysts prior to CVD growth to produce a wide range of CNT thin film densities. Standard photolithography was observed to be incompatible with ferric nitrate processing. Positive photoresist (OIR 908-35, Futurrex Inc) was found to react with hexane solution used for precipitation of ferric nitrate. Lift-off of negative photoresist (NR7 1500 PY, Futurrex Inc) also affected the catalyst unfavorably. This limitation was successfully overcome by using PMMA as a lift-off layer in a triple-layered resist approach. Negative photoresist pattern acts as a mask to etch PMMA in $O_2$ plasma, whereas, the underlying photoresist mesa avoids direct exposure of the growth substrate with $O_2$ plasma that is unfavorable to transfer printing. Thus, this approach ensures compatibility with catalyst processing as well as transfer printing. An SEM image of a patterned CNT thin film in Fig. 2(b) shows that edge roughness is limited by CNT lengths. The density of CNTs shown is high (1.1 CNTs/$\mu m^2$) yet the edge resolution is relatively good. As mentioned earlier, the CNT thin film is pristine. However, a band of greater density of CNTs is observed near the edge, possibly due to increased deposition of catalyst near the edge of the photoresist due to wetting of the photoresist/substrate boundary.

In addition, the patterning can be successfully done post-growth by transfer printing. As illustrated in Fig. 3, the PET stamp is used to subtract CNTs from the growth substrate. Edge roughness of patterned CNT thin films is expected to depend on the stamp material, density, and length of the CNTs. The edge roughness of ± 5 μm was obtained for CNT density = 0.57 CNTs/$\mu m^2$, Fig. 3(b). Since the CNT TFTs have channel area larger than length of individual CNTs the edge roughness of ± 5 μm may be acceptable for large area electronic applications. We have successfully achieved 50 μm x 50 μm to 500 μm x 500 μm square patterns of CNTs using 16 μm deep PET stamps. However, the patternable area is limited by depth of PET stamps due to the side wall collapsing during printing. It was observed that the depth of PET trench decreased from 16 μm to 3 μm after one printing step. Typically a PET stamp can be used effective only once.

To overcome the inherent difficulties of patterning with PET stamps, patterning with Au stamps was investigated (Fig. 3(c)). Fig. 3(d) illustrates that Au stamps were successfully used to pattern the CNT thin films. Clearly, a parallel array of CNT thin film can be produced. Patterned arrays have been made with features as small as 5 μm and edge roughness less than 1 μm. Using this post-growth scheme to pattern as-grown CNT thin films, resolution comparable to that of photolithography can be achieved (Fig. 1(c)). However, the uniformity in stamp thickness and cleanliness of Au are critical for reliable patterning over large areas. We find that flexible stamps (PET) are appropriate to make low resolution patterns over large area, whereas rigid Au stamps are preferred to make high resolution patterns over small area.

Transfer printing was used to pattern as-grown CNT thin films as well as to incorporate them into bottom gate CNT TFTs. Required differential adhesion was achieved via micro-contact printing of a release layer. Using this approach, we achieved almost complete (except few short CNTs) transfer of as-grown patterned CNTs onto PMMA, an improvement over previous reported devices [10]. In ref [10], CNT thin films were patterned by standard post-growth pattern method using RIE resulting in a partial transfer of CNTs onto PMMA (and even back-transfer of PMMA from device substrate to CNT/$SiO_2$ transfer substrate). This could be possibly due to inadequate differential adhesion resulting from presence of a thin photoresist layer left on CNTs after the lift-off [13].

A schematic of final bottom gate device geometry is shown in the inset in Fig. 7(a). Fig. 7(b) shows an SEM image of transfer printed CNTs onto PMMA/$Al_2O_3$ dielectric bi-layer. Output and transfer characteristics of a device ($L = 70$ μm) are shown in Fig. 7(c) and (d), respectively. The density of CNTs in this device was 0.57 CNTs/μm$^2$. Field-effect mobility of this device is 12.5 cm$^2$/Vs and on/off ratio of $3 \times 10^3$. Bottom gate CNT TFTs show *p*-type behavior with gate-bias hysteresis, consistent with bottom-gate devices reported earlier [10]. The field-effect mobility of devices was observed to be in range $2 - 20$ cm$^2$/Vs with on/off ratio in range $200 - 4 \times 10^3$. In comparison, RIE-patterned CNT thin film devices had shown field-effect mobility in range 0.08 – 13.7 cm$^2$/Vs with on/off ratio between $10^2$ – $10^3$. The low field-effect mobility of RIE patterned bottom gate CNT devices could be due to partial transfer and/or poor contacts of resist-exposed CNTs on the device substrate. [The field-effect mobility ($\mu$) is calculated as $\mu = \dfrac{L}{V_d C_g} \left| \dfrac{\partial I_d}{\partial V_g} \right|$, where $C_g$ is gate-capacitance, $C_g = \dfrac{\varepsilon LW}{d}$, where $W$ is channel width, $L$ is channel length, $V_d$ is drain bias, $I_d$ is drain current, $V_g$ is gate bias, $d$ is gate dielectric thickness and $\varepsilon$ is dielectric contact of the gate dielectric. $C_g$ is calculated by assuming a continuous semiconducting film in the device channel instead of partial coverage of CNT network (< 0.1%), and therefore, calculated field-effect mobility is an underestimation of true field-effect mobility of CNTs. The field-effect mobility was calculated from reverse $I_d$ – $V_g$ sweeps.]

**4. Discussion**

The methods presented here enable fabrication of high quality CNT thin films for arrays of devices. The density of CNT thin films was controlled by changing gas flow rate and concentration of ferric nitrate catalyst. The first method provided fine control over density of sparse network, whereas, the second method produced CNT density over a wider range, 0.04 CNTs/μm$^2$ – 1.29 CNTs/μm$^2$. These CNTs thin films of varying density enable the optimization for high performance CNT devices. The effect of growth temperature on CNT density is beyond the scope of this work. However, the growth temperature in the range 800 – 900 °C is expected to produce optimum yield of single-walled nanotubes for catalysts used in this work [7, 14, 15]. Density-dependent percolation effects have also been investigated, but it is outside the scope of this work and will be reported elsewhere.

The pristine thin films of CVD-grown CNTs can be successfully patterned through pre-growth as well as post-growth methods. Usual pre-growth patterning methods were expanded to make them compatible with ferric nitrate processing and transfer printing. Furthermore, post-growth patterning schemes can produce patterned as-grown CNT thin films with edge roughness (< 1 μm) and resolution (< 5 μm) that are comparable to that from RIE. Relative advantages (and disadvantages) of different patterning methods are shown in Table 1. The first two methods (pre-growth) produce pristine CNTs, but they are limited by edge-roughness of CNT patterns due to CNT growth exterior to the patterned catalyst. One possible way to achieve better edge roughness is to inhibit CNT growth in the exterior. At present, we are unaware of any technique that can achieve that. The last two methods (post-growth) produce comparatively better edge roughness (< 1 μm). However, transfer printing method is preferred for pristine CNT thin films. All the patterning methods were successfully incorporated into transfer printing by treating the growth substrate with release layer through micro-contact printing, which could be generalized other patterning methods not reported here.

The resulting CNT thin film have been successfully incorporated in bottom gate TFTs on PET substrates using PMMA/$Al_2O_3$ bi-layer as gate dielectric and Au as source, drain and gate electrodes. There are three possible advantages of the current approach of device assembly over previously reported resist-processed CNT devices [10]. First, a complete transfer printing of CNTs in current devices ensures better control of the density of CNTs in the device channel. Second, absence of residual resist could be used to make better contact with electrodes. Third, the diffusion barrier of $Al_2O_3$ film allows reduced gate leakage and higher gate capacitance. Finally, transfer printing methods have also been used elsewhere to demonstrate CNT thin film electrodes and organic semiconductor based transparent, flexible electronics [16]. An additional advantage of transfer printing approach is realized by fabrication of as-grown suspended CNT devices to study intrinsic properties of CNTs [12].

## 5. Conclusions

We have demonstrated methods that successfully allow controlled growth, patterning, and placement of CNT thin film for electronic applications. Controlled density of CNTs was obtained by optimizing the gas flow-rate as well as the concentration of the growth catalyst in a CVD process. The CNT thin films were then successfully patterned using both pre-growth and post-growth patterning techniques. Significant improvements over typical pre-growth patterning methods were made by developing layered photoresist process that is compatible with ferric nitrate. Limitations of traditional post-growth patterning methods, such as exposure to resist processing, were circumvented by transfer printing CNTs selectively from the growth substrate to either thermoplastic (PET)) or metallic (Au) stamps. Resulting CNT thin film patterns are pristine, and have high resolution (< 5 μm) and low edge roughness (< 1 μm). Finally, transfer printing methods were used to achieve controlled placement of patterned CNT thin films on different plastic substrates. Bottom gate TFTs showed field-effect mobility up to 20 cm$^2$/Vs with on/off ratio of the order of $10^3$. The methods presented here could be possibly generalized to make suspended devices with other materials, such as nanowires and graphene, to study their intrinsic properties.

**Acknowledgments:** This work has been supported by the Laboratory for Physical Sciences and by the use of the UMD-MRSEC SEF Grant No. DMR 05-20471. Infrastructure support is also provided by the UMD NanoCenter and CNAM.

**Figures:**

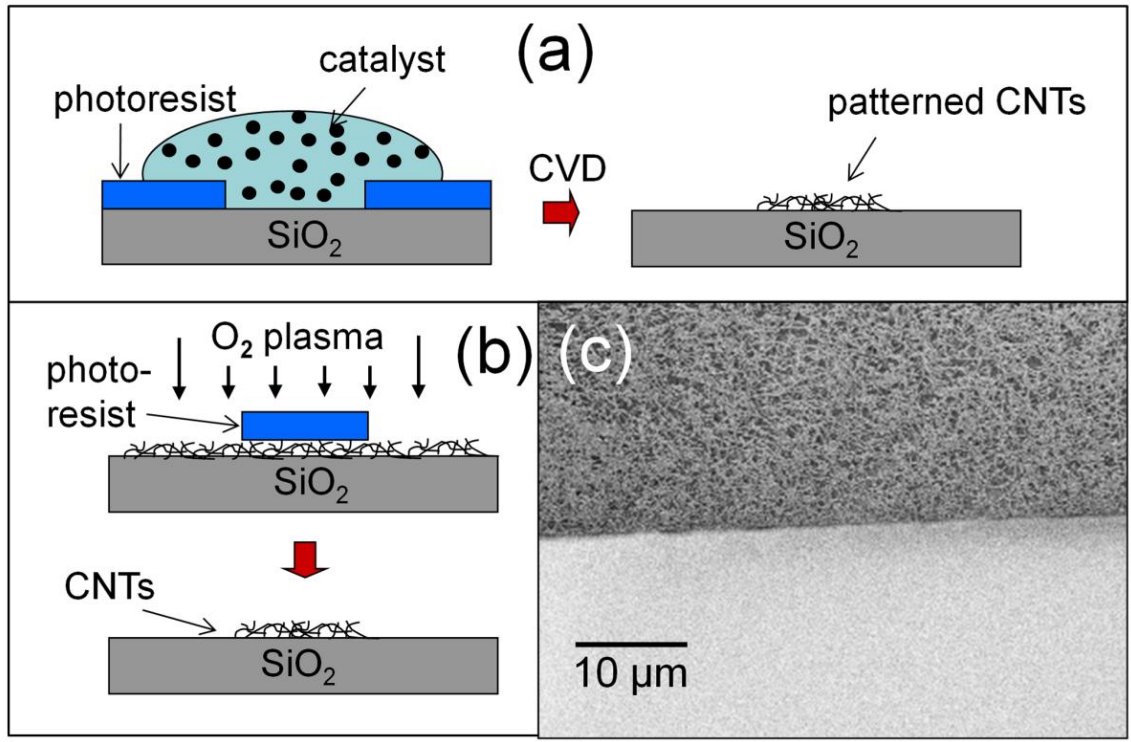

Figure 1. a) Schematic of the steps used in a standard pre-growth patterning method. A catalyst is patterned by photolithography and CNTs grown by CVD after the lift-off. b) Schematic of the steps used in a standard post-growth patterning method. CNT thin film patterned using photolithography and reactive ion etching. c) FE-SEM micrograph of the edge of a CNT thin film patterned by post-growth method in (b). This CNT thin film was grown by the second growth method with a catalyst concentration of 133.3 μg/ml.

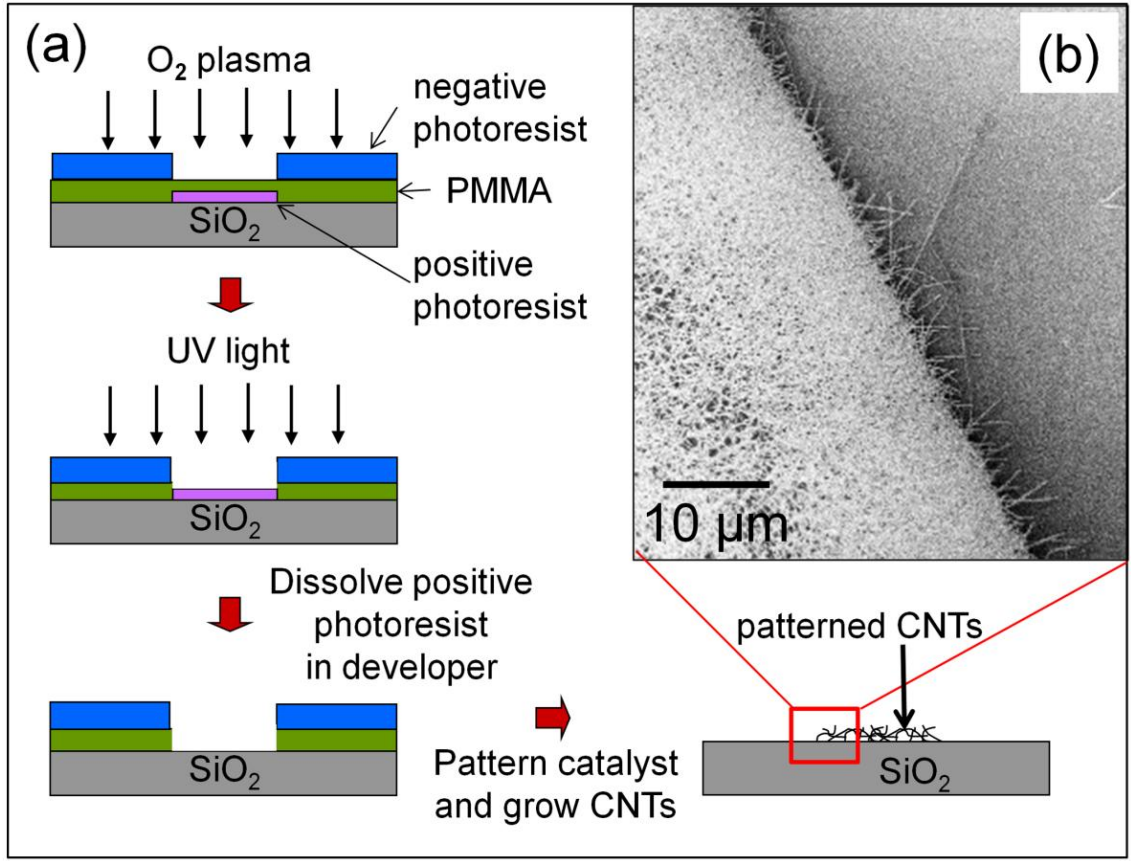

Figure 2. a) Schematic of the photolithography steps used to synthesize patterned CNT thin films using pre-patterning of the catalyst via a bilayer of photoresist and PMMA. The direct exposure of $SiO_2$ to $O_2$ plasma is avoided by an extra layer of positive photoresist underneath PMMA. b) FE-SEM image of the edge of a patterned CNT thin film grown by the second growth method with a catalyst concentration of 66.7 μg/ml.

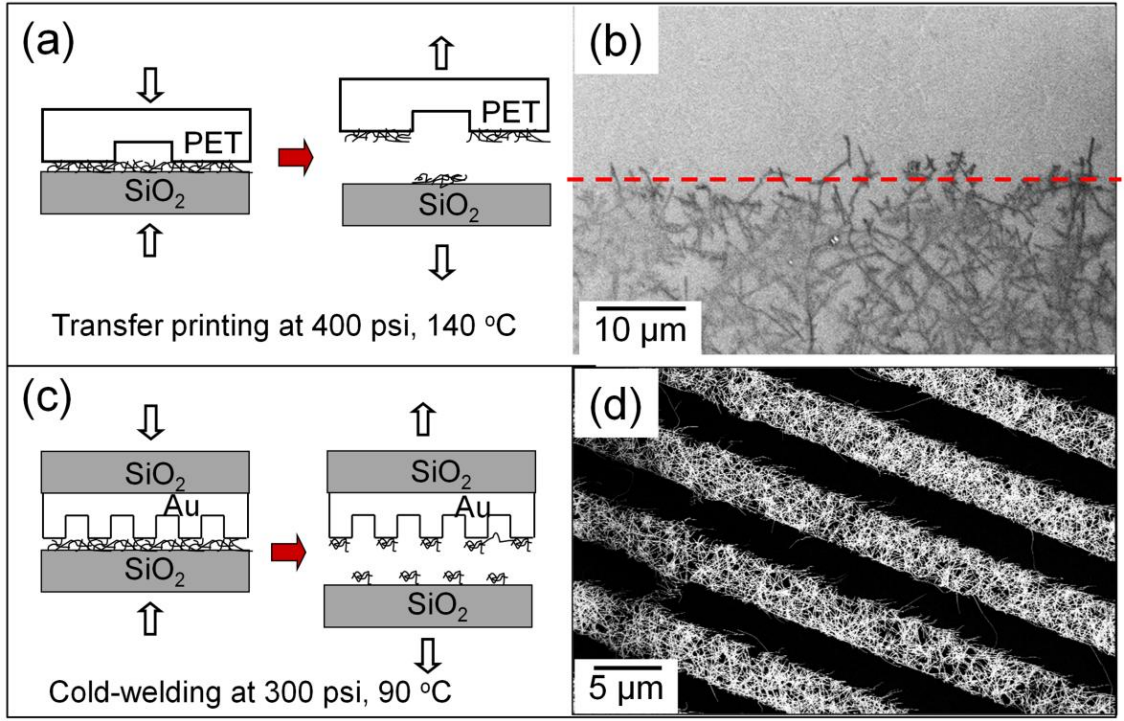

Figure 3. a) Schematic of patterning CNT thin films by transfer printing. CNT thin film is pressed against PET stamp at 400 psi, 140 °C for 3 min. b) FE-SEM image of the edge of a CNT thin film patterned by transfer printing. The CNT thin film was grown by using the second growth method with a catalyst concentration of 13.3 μg/ml. c) Schematic of patterning CNT thin films by cold-welding CNTs to a Au stamp at 300 psi, 90 °C for 3 min. d) FE-SEM image of strips of CNT thin film patterned by Au stamps. The CNT thin film was grown by using the second growth method with a catalyst concentration of 133.3 μg/ml.

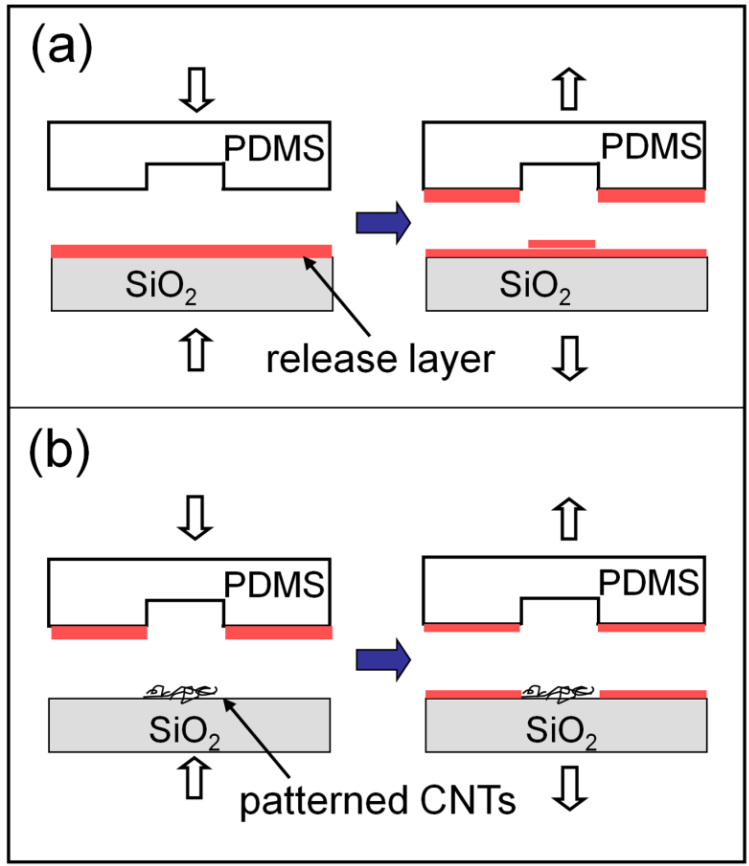

Figure 4. a) Schematic of treating the raised features of a PDMS stamps with release layer from release layer-coated Si substrate by micro-contact printing. b) Release layer-coated PDMS stamps are then used to pattern release layer on CNT-covered $SiO_2$ surface. Release layer is printed only in the areas exterior to the patterned CNT thin films. Vertical arrows represent that the substrates are brought in to and out of contact gently. No pressure was applied on PDMS stamps.

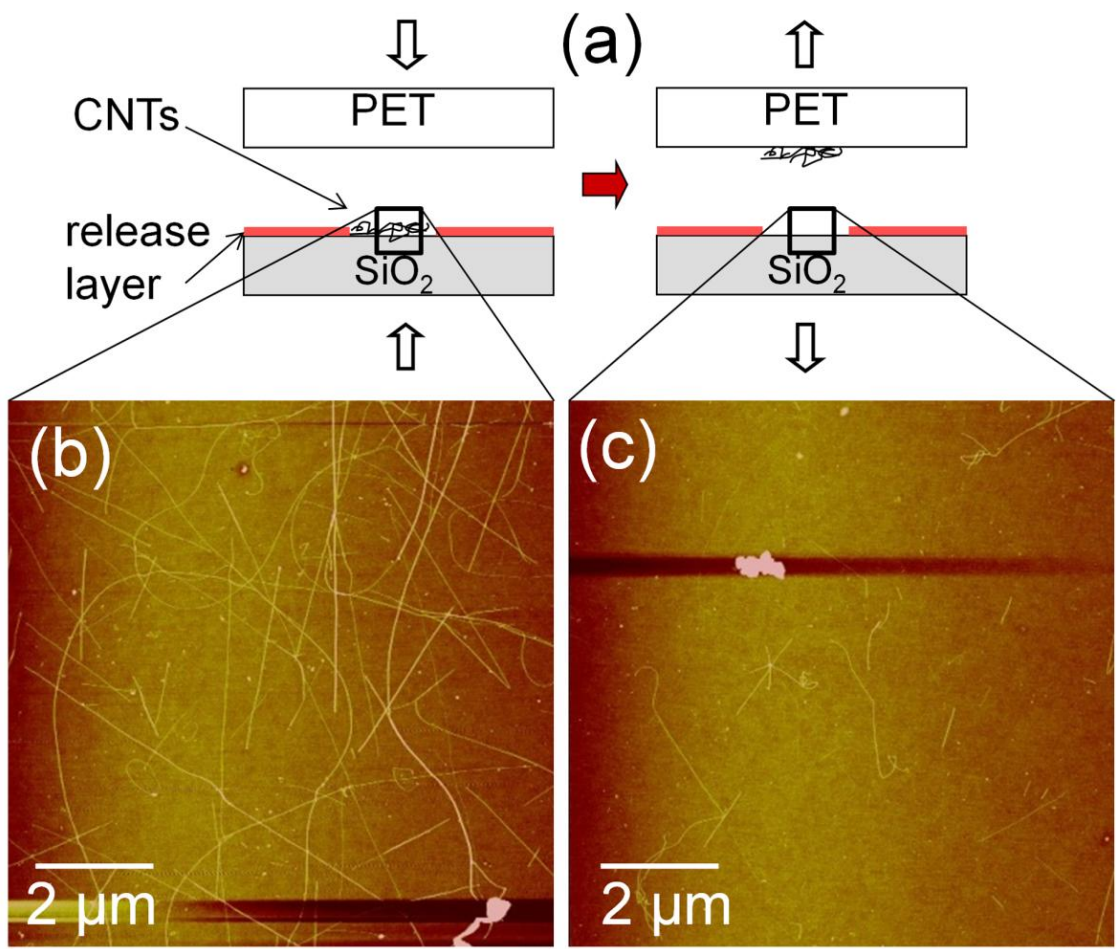

Figure 5. a) Transfer printing CNTs from release-layer treated SiO$_2$/Si substrate to a PET substrate at 500 psi, 170 $^{o}$C for 3 min. b) AFM images of the transfer substrate (SiO$_2$/Si) before transfer printing. CNTs were grown using the second growth method with a catalyst concentration of 33.3 μg/ml. c) AFM image of the transfer substrate after transfer printing.

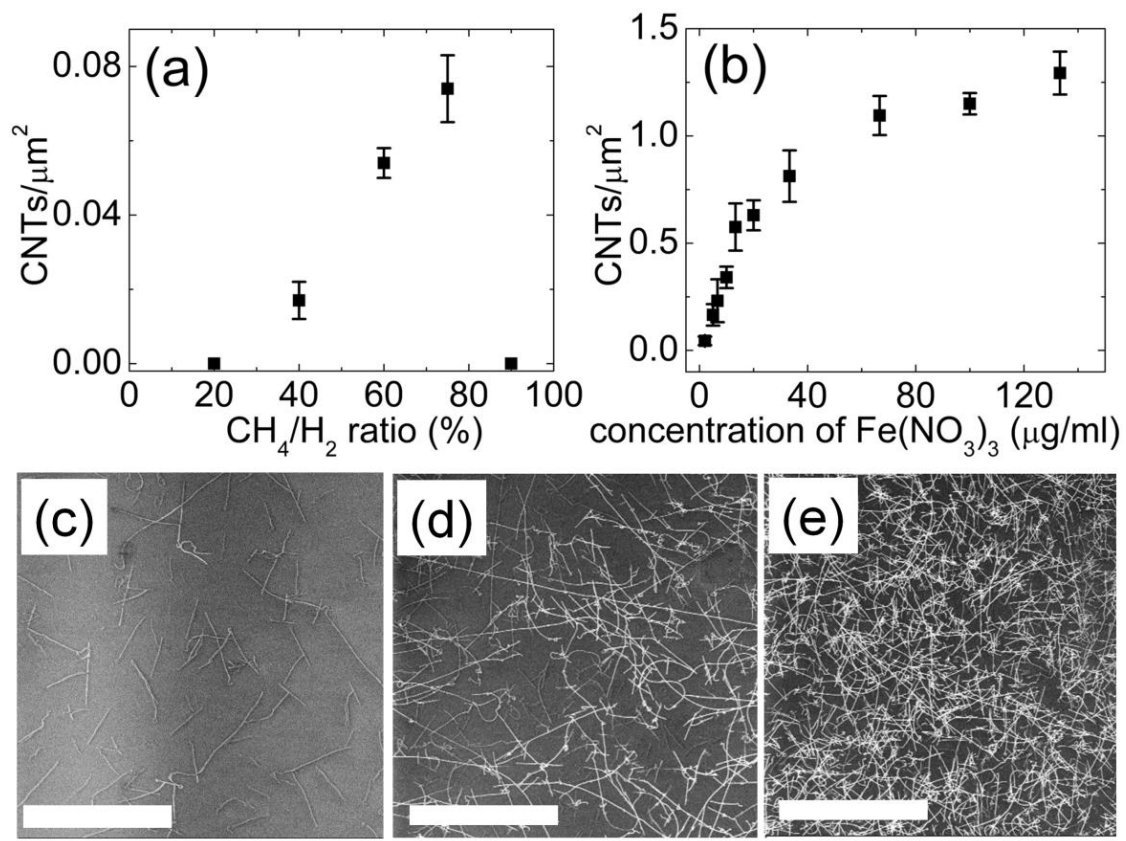

Figure 6. a) CNT network density as a function of ratio of CH$_4$ to total gas feed. b) CNT network density is plotted against the concentration of ferric nitrate solution. c), d) and e) SEM images of CNT thin films grown by the second method (b) with CNT densities 0.16 CNTs/μm$^2$, 0.63 CNTs/μm$^2$ and 1.15 CNTs/μm$^2$, respectively. The scale bar is 10 μm in all the images.

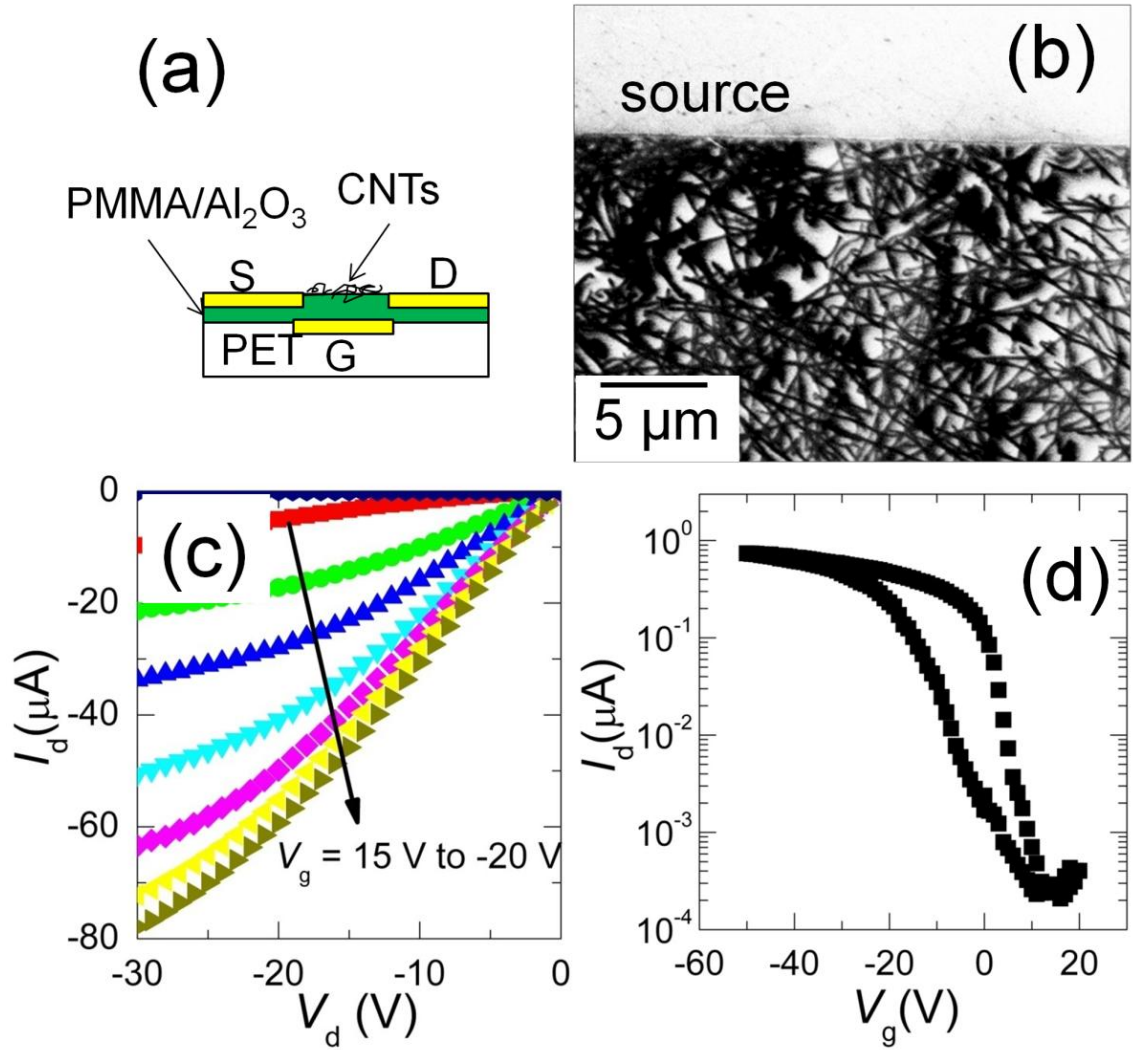

Figure 7. a) Schematic of the bottom gate device assembled on a PET substrate. b) SEM image of CNTs in the devices channel printed onto PMMA dielectric layer. b) Output characteristics of a CNT TFT with $L = 70$ μm, $W = 100$ μm. $I_d - V_d$ curves are taken at $V_g = 15$ V to $-20$ V in steps of 5 V. d) Transfer curve of the device at $V_d = -1$ V.

| | Patterning methods | Advantage | Disadvantage | Transfer printing | Edge roughness |
|---|---|---|---|---|---|
| Pre-growth | 1. Patterning catalyst by photoresist (Fig.1(a)) | • No chemical exposure | • Incompatible with $Fe(NO_3)_3$ | Yes | Limited by CNT length (~ 10 μm) |
| Pre-growth | 2. Patterning catalyst by triple-layered resist (Fig.2) | • No chemical exposure<br>• Compatible with $Fe(NO_3)_3$ | • Cumbersome | Yes | Limited by CNT length (~ 10 μm) |
| Post-growth | 3. Reactive ion etching (Fig. 1(b)) | • Straightforward | • Exposure to chemicals | Yes | Limited by photo-lithography (< 1 μm) |
| Post-growth | 4. Transfer printing (Fig. 3,4 and 5) | • No chemical exposure<br>• Compatible with $Fe(NO_3)_3$ | •Additional fabrication of stamps and micro-contact printing | Yes | Limited by stamps (~ 10 μm (PET))<br>(~ 1 μm (Au)) |

Table 1. A comparison of various methods to pattern CNT thin films. Method 1 and 3 are popular pre-growth and post-growth patterning methods. Method 2 and 4 are developed in this work.

**Reference:**

1. Kumar, S., J.Y. Murthy, and M.A. Alam, *Percolating conduction in finite nanotube networks.* Physical Review Letters, 2005. **95**.
2. Snow, E.S., et al., *Random networks of carbon nanotubes as an electronic material.* Applied Physics Letters, 2003. **82**(13): p. 2145-2147.
3. Unalan, H.E., et al., *Design criteria for transparent single-wall carbon nanotube thin-film transistors.* Nano Letters, 2006. **6**(4): p. 677-682.
4. Franklin, N.R., et al., *Patterned growth of single-walled carbon nanotubes on full 4-inch wafers.* Applied Physics Letters, 2001. **79**(27): p. 4571-4573.
5. Tselev, A., et al., *A photolithographic process for fabrication of devices with isolated single-walled carbon nanotubes.* Nanotechnology, 2004. **15**(11): p. 1475-1478.
6. Kong, J., et al., *Synthesis of individual single-walled carbon nanotubes on patterned silicon wafers.* Nature, 1998. **395**(6705): p. 878-881.
7. Edgeworth, J.P., N.R. Wilson, and J.V. Macpherson, *Controlled growth and characterization of two-dimensional single-walled carbon-nanotube networks for electrical applications.* Small, 2007. **3**(5): p. 860-870.
8. Huang, S.M., et al., *Patterned growth of well-aligned carbon nanotubes: A soft-lithographic approach.* Journal Of Physical Chemistry B, 2000. **104**(10): p. 2193-2196.
9. Cao, Q., et al., *Medium-scale carbon nanotube thin-film integrated circuits on flexible plastic substrates.* Nature, 2008. **454**(7203): p. 495-U4.
10. Sangwan, V.K., et al., *Patterned Carbon Nanotube Thin-Film Transistors with Transfer-Print Assembly.* Mater. Res. Soc. Symp. Proc., 2007. **963**: p. 0963-Q0910-0957.
11. Hines, D.R., et al., *Nanotransfer printing of organic and carbon nanotube thin-film transistors on plastic substrates.* Applied Physics Letters, 2005. **86**(16).
12. Sangwan, V.K., et al., *Facile fabrication of suspended as-grown carbon nanotube devices.* Applied Physics Letters, 2008. **93**(11).
13. Hines, D.R., et al., *Transfer printing methods for the fabrication of flexible organic electronics.* Journal Of Applied Physics, 2007. **101**(2).
14. Kim, W., et al., *Synthesis of ultralong and high percentage of semiconducting single-walled carbon nanotubes.* Nano Letters, 2002. **2**(7): p. 703-708.
15. Li, Y.M., et al., *Growth of single-walled carbon nanotubes from discrete catalytic nanoparticles of various sizes.* Journal Of Physical Chemistry B, 2001. **105**(46): p. 11424-11431.
16. Southard, A., et al., *Solution-processed single walled carbon nanotube electrodes for organic thin-film transistors.* Organic Electronics, 2009. **10**(8): p. 1556-1561.